\documentclass[%
 reprint,
superscriptaddress,
 amsmath,amssymb,
 aps,
pra,
]{revtex4-2}

\usepackage{amsmath}
\usepackage{amssymb}
\usepackage{amsthm}
\usepackage{physics}
\usepackage{graphicx}
\usepackage{dcolumn}
\usepackage{bm}
\usepackage{hyperref}

\theoremstyle{definition}
\newtheorem{thm}{Theorem}

\newtheorem{dfn}[thm]{Definition}

\usepackage{xcolor}

\begin{document}

\preprint{APS/123-QED}

\title{Long-Distance Device-Independent Conference Key Agreement}

\author{Makoto Ishihara}
 \email{llmakomako.arg1076@keio.jp}
 \affiliation{%
 Department of Electronics and Electrical Engineering, Keio University, 3-14-1 Hiyoshi, Kohoku-ku, Yokohama 223-8522, Japan
}%
\author{Anders J. E. Bjerrum}
\affiliation{%
 Center for Macroscopic Quantum States (bigQ), Department of Physics, Technical University of Denmark, 2800 Kongens Lyngby, Denmark
}%
\author{Wojciech Roga}%
 \affiliation{%
 Department of Electronics and Electrical Engineering, Keio University, 3-14-1 Hiyoshi, Kohoku-ku, Yokohama 223-8522, Japan
}%
\author{Jonatan B. Brask}
\affiliation{%
 Center for Macroscopic Quantum States (bigQ), Department of Physics, Technical University of Denmark, 2800 Kongens Lyngby, Denmark
}%
\author{Ulrik L. Andersen}
\affiliation{%
 Center for Macroscopic Quantum States (bigQ), Department of Physics, Technical University of Denmark, 2800 Kongens Lyngby, Denmark
}%
\author{Masahiro Takeoka}%
\affiliation{%
 Department of Electronics and Electrical Engineering, Keio University, 3-14-1 Hiyoshi, Kohoku-ku, Yokohama 223-8522, Japan
}%
\affiliation{
National Institute of Information and Communications Technology (NICT), Koganei, Tokyo 184-8795, Japan.
}

\date{\today}

\begin{abstract}
Device-independent quantum key distribution (DI-QKD) enables two remote parties to share an information-theoretically secure key without any assumptions on the inner workings of the devices used. Device-independent conference key agreement (DI-CKA) is multipartite DI-QKD where more than two parties share a common secure key. The performance of DI-CKA, however, is strictly limited because of its susceptibility to losses due e.g. to imperfect detection efficiency and channel transmission. Here, we propose a DI-CKA protocol which reduces this limitation by using a heralding scheme to distribute multipartite entanglement. We analyze key rates of our protocol for two different measurement scenarios and we show that our protocol outperforms a previous DI-CKA protocol even with an experimentally feasible measurement.

\end{abstract}

\maketitle


\section{\label{sec:introduction}INTRODUCTION}
Quantum key distribution (QKD)~\cite{Bennett2014, Ekert1991, Xu2020, Pirandola2020} enables two legitimate parties to share an information-theoretically secure key. In standard QKD protocols, security is guaranteed based on thorough characterization of the devices used to implement the protocols. However, if an eavesdropper, Eve, performs side-channel attacks not accounted for this characterization, security can be compromised. Therefore, patching for identified side-channel attacks is used in real QKD demonstrations \cite{Xu2020}. Such patching strategies, however, cannot completely remove a threat of compromising security since unknown side-channel attacks may exist.

Device-independent QKD (DI-QKD) \cite{Acin2007, Pironio2009} can be secure without the above assumption, that is, it does not require any assumptions on the inner workings of the devices. A lot of theoretical research on DI-QKD has been carried out, see Refs.~\cite{Zapatero2023, Primaatmaja2023}. Moreover, recently, proof-of-principle experimental demonstrations have been performed~\cite{Nadlinger2022, Zhang2022, Liu2022}. 

While in DI-QKD protocols a common secret key is shared between two legitimate parties, in device-independent conference key agreement (DI-CKA), the number of legitimate participants who share the common key is $N\geq 3$. The underlying principle behind typical DI solutions is checking violation of some Bell inequalities by measuring shared quantum states. Violation of Bell inequalities constrains the possible states and measurements that the devices could use, and it limits the maximum amount of information that could potentially leak to an eavesdropper. So far, several DI-CKA protocols based on the Greenberger-Horne-Zeilinger (GHZ) states and multipartite Bell inequalities have been proposed~\cite{Ribeiro2018, Holz2019, Ribeiro2019, Holz2020,Grasselli2023}. However, the range over which these protocols can distribute a secret key is very limited as the key rate is highly susceptible to implementation imperfections related to, for instance, limited channel transmission rate and limited photodetection efficiency. The imperfections reduce applicability of these DI-CKA protocols in practical QKD network.

In this paper, we propose a DI-CKA protocol which can distribute a secret key over long distances of tens of kilometers. To overcome the loss caused by channel transmission, we utilize the heralding scheme to efficiently distribute a GHZ state proposed in~\cite{Shimizu2025}. Such heralding schemes have been widely used in DI-QKD protocols~\cite{Gisin2010, Zapatero2019, Kolodynski2020}. We consider two scenarios for the measurement performed by each party. In one scenario, each party can perform arbitrary qubit measurements. Implementing such an arbitrary measurement is, however, not straightforward. For instance, for states encoded in the Fock basis, a measurement of $\sigma_X$ requires projection onto superpositions of different photon-number states. While such measurements can be realized using either matter-based quantum memories~\cite{Ashhab2007, Nadlinger2022, Zhang2022} or Dolinar-inspired detectors~\cite{Takeoka2005, Takeoka2006}, this is still technologically challenging. Therefore, we consider also a second scenario where each party instead performs displaced single-photon detection (i.e. measurements combining optical displacement operations and photodetection). An advantage of such measurements is that their experimental realization is feasible with current technology. For the former scenario, we calculate a key rate of our protocol based on violation of the parity-CHSH inequality~\cite{Ribeiro2018}. For the latter, we calculate a key rate by using a
numerical optimization approach~\cite{Navascues2008, Masanes2011, NietoSilleras2014, Bancal2014, Brown2024}. We show that our protocol can distribute a secret key over longer distances than the previous protocol \cite{Ribeiro2018, Holz2019, Ribeiro2019} for both scenarios.

The paper is structured as follows. In Sec. \ref{section:protocol} we describe our long-distance DI-CKA protocol. We explain two scenarios of measurement in Sec. \ref{section:measurementscenario}. Then, we show how to calculate key rates of our protocol in Sec. \ref{section:keyrate}. In Sec. \ref{section:results} we provide results of the key rate calculation, and finally we conclude in Sec. \ref{section:conclusion}.

\section{PROTOCOL}\label{section:protocol}
\begin{figure}[tbp]
\includegraphics[keepaspectratio, scale=0.38]{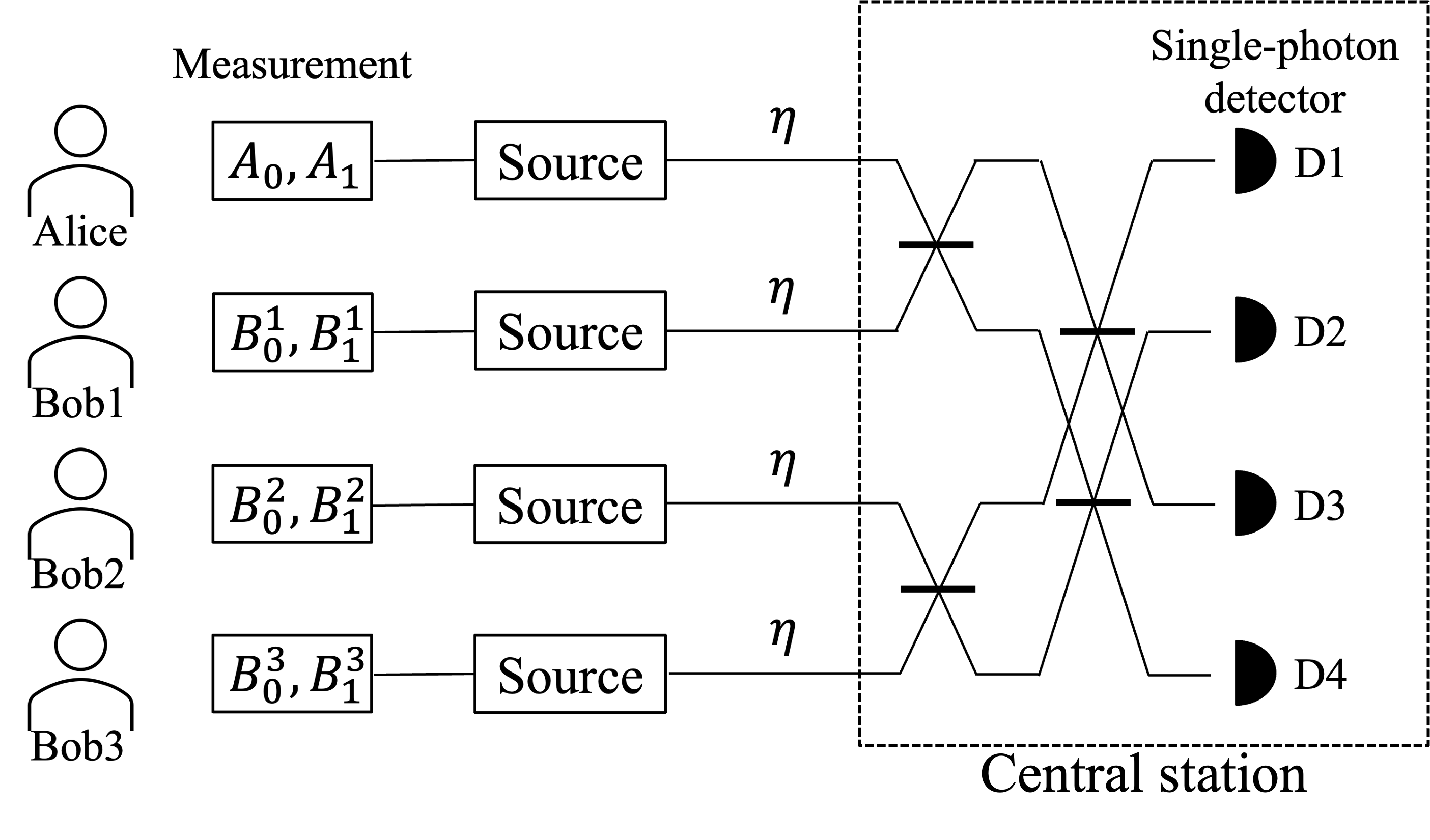}
\caption{\label{fig:DICKASetup4}
Schematic of our DI-CKA protocol for $N=4$. Each party prepares a single-photon entanglement $\sqrt{q}\ket{10} + \sqrt{1-q} \ket{01}$ with a single-photon source. They transmit one part of the single-photon entanglement to the central station through pure-loss channels with transmissivity $\eta$. At the central station, these modes are interfered with 50:50 beamsplitters and detected with single-photon detectors. They use events where the single-photon detectors $D_1$ and $D_2$ detect a single-photon to share a common secret key.}
\end{figure}

In this section, we describe our DI-CKA protocol. Consider that $N$ legitimate parties: Alice and $\text{Bob}_{i}$, where $i \in \{ 1, \ldots , N-1 \}$, try to share a common secret key. We assume that $N$ is an even number since the efficient GHZ distribution scheme cannot be applied to an odd number of parties \cite{Shimizu2025}. At first, we focus on the situation where $N=4$, as is schematically represented in Fig.~\ref{fig:DICKASetup4}. For every round, each party prepares a single-photon entangled state of two-modes, $X$ and $X'$,
\begin{equation}
\sqrt{q} \ket{10}_{XX'} + \sqrt{1-q} \ket{01}_{XX'},
\label{initial}
\end{equation} 
where mode $X'$ is transmitted to a central station through a pure-loss channel with transmissivity $\eta$ and mode $X$ is detected by the party. At the central station, these four modes are interfered in an interferometer consisting of 50:50 beamsplitters, as shown in Fig.~\ref{fig:DICKASetup4}, and detected with single-photon detectors, where each single-photon detector has finite detection efficiency $\eta_d$ and dark count probability $p_{DC}$. When only two of the four single-photon detectors detect a single-photon, the state shown in Tab.~\ref{Tab:DistributedGHZstate} is distributed over the four parties. We call such an event a successful event. In our protocol, we focus on the event where the detectors $D_1$ and $D_2$ detect a single-photon but the same analysis holds for the other click patterns. Details regarding the interferometer at the central station can be found in Appendix~\ref{appendixA}. 


After successful generation of a GHZ state every party performs some measurements on the distributed state. In practice, the parties perform measurements regardless of whether successful events occur or not, and they discard events where the heralding does not succeed afterward. Each successful event can be classified into two types. One is key generation round and the other one is Bell test round. For the key generation round, every party performs $\sigma_{Z}$ measurement on the GHZ state. For the Bell test round, each party performs one of two measurements with binary outcomes, that is, Alice performs $A_x$, where $x \in \{0, 1 \}$ is Alice's measurement input, and $\text{Bob}_{i}$ performs $B_{y_i}^{i}$, where $y_i \in \{0, 1\}$ is $\text{Bob}_i$'s measurement input. Let $a \in \{0, 1\}$ denote Alice's output and $b_i \in \{0,1\}$ denote $\text{Bob}_i$'s output. Here, we assume that the choices of measurement inputs are causally disconnected from the heralding results at the central station \cite{Kolodynski2020}. Also, we assume that each party's measurement has finite detection efficiency $\eta_e$ and dark count probability $p_{DC}^e$. After the measurement rounds, the legitimate users perform parameter estimation and calculate a DI-CKA key rate. If their calculation shows a non-zero key rate, they perform error correction and privacy amplification. Otherwise, they abort the protocol. We can extent this protocol to more than four parties in the following way, see also Appendix~\ref{appendixA}. For instance, when $N=6$, we prepare two copies of the four-party-interferometer and input a single-photon entangled state into two edge modes~\cite{Shimizu2025}. Extension to more than 6 parties is a straightforward generalization of this construction.

\begin{table}[tbp]
\begin{center}
   \caption{Distributed GHZ states depending on the click patterns of the detectors at the central station. $\ket{\psi}$ is a quantum state distributed over legitimate parties when each click pattern happens.}
   \label{Tab:DistributedGHZstate}
  \begin{tabular}{c|c} \hline
     Detectors which react & $\ket{\psi}$   \\ \hline
    $D_1, D_2$ & $(\ket{0101}-\ket{1010})/\sqrt{2}$ \\ \hline
    $D_1, D_3$ & $(\ket{0011}-\ket{1100})/\sqrt{2}$  \\ \hline
    $D_1, D_4$ & $(\ket{0110}-\ket{1001})/\sqrt{2}$  \\ \hline
    $D_2, D_3$ & $(\ket{1001}-\ket{0110})/\sqrt{2}$   \\ \hline
    $D_2, D_4$ & $(\ket{1100}-\ket{0011})/\sqrt{2}$   \\ \hline
    $D_3, D_4$ & $(\ket{1010}-\ket{0101})/\sqrt{2}$   \\ \hline 
  \end{tabular}
\end{center}
\end{table}

\section{TWO MEASUREMENT SCENARIOS}\label{section:measurementscenario}
\begin{figure}[tbp]
 \centering
 \includegraphics[keepaspectratio, scale=0.45]{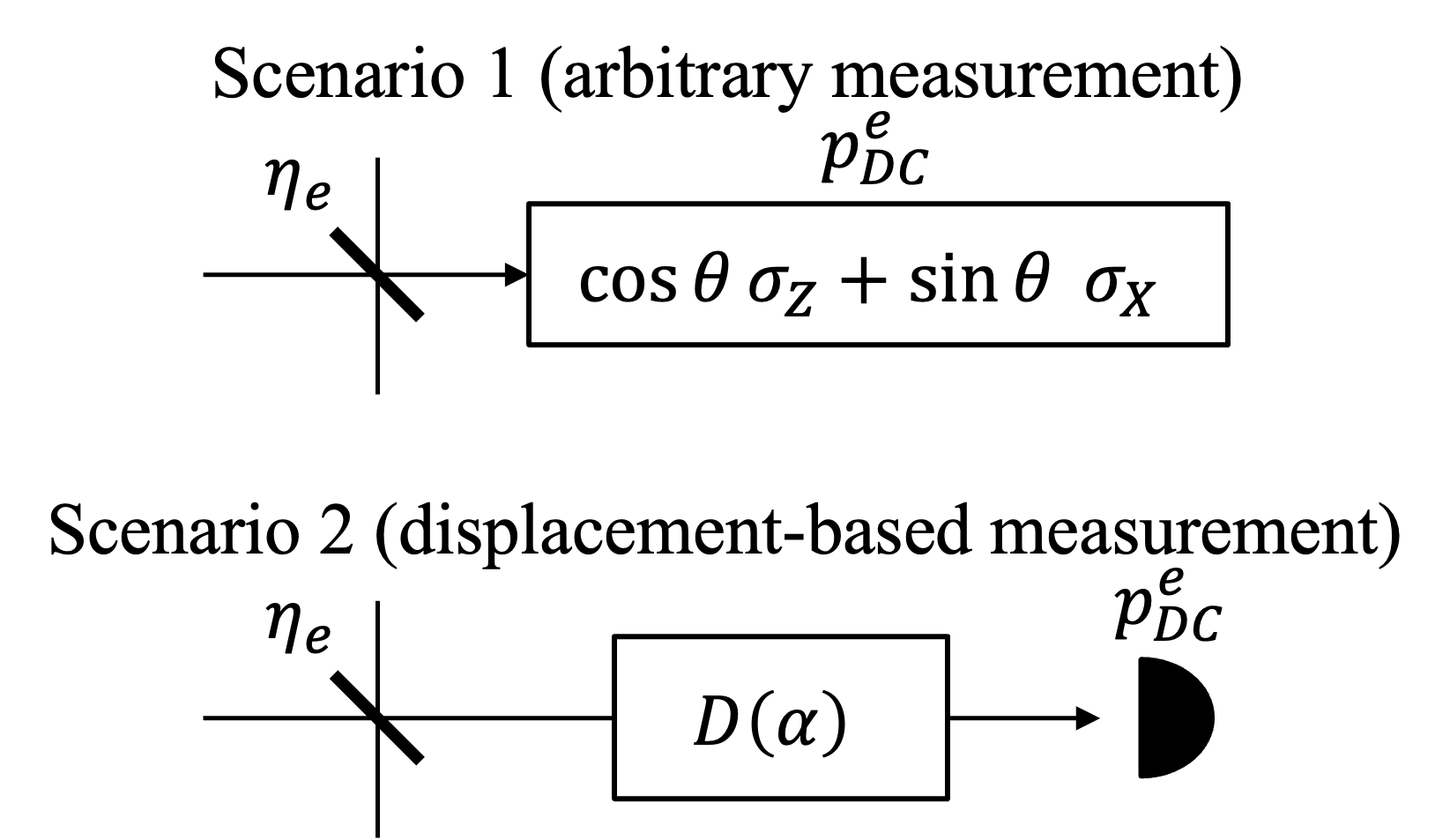}
 \caption{Schematics of two measurement scenarios. In Scenario 1, we model the detection efficiency $\eta_e$ as a pure-loss channel with transmissivity $\eta_e$ followed by measurements in the $X$-$Z$ plane of the Bloch sphere with unit efficiency and dark count probability $p_{DC}^e$. In Scenario 2, we effectively shift the detection efficiency of the photon detector prior to the displacement operation. Again, we model the detection efficiency $\eta_e$ as a pure-loss channel.}
 \label{fig:Measurement}
\end{figure}

Here, we describe details regarding the measurements performed by the legitimate parties. In this paper, we consider two measurement scenarios: we call the first one Scenario 1 and the second Scenario 2. Schematics of these measurement scenarios are shown in Fig. \ref{fig:Measurement}. In Scenario 1, every party can perform an arbitrary Pauli measurement, such as $\sigma_Z$ and $\sigma_X$. However, since the distributed GHZ states are encoded in the photon-number basis, experimental implementation of such arbitrary measurement is not straightforward. Although implementation of such measurement is not easy, we can realize it by using matter-based quantum memories~\cite{Ashhab2007, Nadlinger2022, Zhang2022} or the Dolinar-inspired detectors~\cite{Takeoka2005, Takeoka2006}. In Scenario 2, every party performs displacement-based measurement, that is, measurement combining displacement operation and photon detection. This measurement allows the parties to perform projections onto coherent states. An advantage of this measurement is that it is implementable with current technology. By considering this measurement, we analyze the feasibility of our protocol in an experiment.

First, we describe the measurement in Scenario 1. Let us define positive operator-valued measures (POVMs) $\Pi (\theta) = \cos{\theta} \sigma_Z + \sin{\theta} \sigma_X$ and $M (\theta) = M_0 (\theta) - M_1 (\theta)$, where
\begin{equation}
    \begin{split}
        M_0 (\theta) &= (1-p_{DC}^e) \frac{I + \Pi (\theta)}{2},\\
        M_1 (\theta) &= I - M_0.
    \end{split}
\end{equation}
Here, these operators do not include the detection efficiency $\eta_e$ since we model this as a pure-loss channel with transmissivity $\eta_e$ followed by measurements with unit efficiency. For the key generation rounds, every party performs the measurement corresponding to $M(0)$. For the Bell test rounds, Alice performs $A_0 = M(0)$ or $A_1 = M (\pi/2)$, $\text{Bob}_1$ performs $B_0^1 = M(-3\pi/4)$ or $B_1^1 = M(3\pi/4)$, and the other Bobs perform $B_0^i = B_1^i = M(\pi/2)$ for $i \in \{2, \ldots , N-1 \}$.

Next, let us specify the measurement in Scenario 2. Measurement with displacement operation and photon detection is defined by the set of measurement operators
\begin{equation}
    \{\ketbra{\alpha},  I-\ketbra{\alpha}  \},
\end{equation}
where $\ket{\alpha}$ is a coherent state
\begin{equation}
    \ket{\alpha} = e^{-\frac{|\alpha|^2}{2}} \sum_{n=0}^{\infty} \frac{\alpha^n}{\sqrt{n!}} \ket{n}.
\end{equation}
Then, considering detection efficiency and the dark count probability, the projectors which each party uses as the measurement operators can be written as
\begin{equation}
    \begin{split}
        M_0 (\alpha) &= (1-p_{DC}^e) \ketbra{\alpha}{\alpha},\\
        M_1 (\alpha) &= I-M_0(\alpha).
    \end{split}
\end{equation}
Here, we effectively shift the efficiency of the detector prior to the displacement operation since we can compensate any loss at the detector by increasing the amplitude. Again, we model the detection efficiency $\eta_e$ as a pure-loss channel with transmissivity $\eta_e$.


Similar to Scenario 1, there are two measurement rounds, that is, the key generation rounds and the Bell test rounds. For the key generation rounds, every party performs $\sigma_Z$ measurement which corresponds to $\alpha = 0$ in this scenario. For the Bell test rounds, every party chooses two values of $\alpha$ and performs one of the two displacement operations on their system. Here, we consider that Alice's displacement is always 0 for $A_0$ measurement and we numerically optimize the other displacements to maximize the key rates.


\section{KEY RATE CALCULATION}\label{section:keyrate}
In this section, we describe how to calculate the key rates of our DI-CKA protocol. We focus on asymptotic key rates and assume that devices behave independently and identically across rounds. A DI-CKA key rate $K$ of our protocol can be expressed as follows \cite{Ribeiro2018}
\begin{equation}\label{eq:keyrate}
    \begin{split}
        K = P_{\text{success}}( H(A|x^*,E) - \max_{i} H(A|B_i,x^*,y_i^*)),
    \end{split}
\end{equation}
where $A$, $B_i$ and $E$ denote Alice's, $\text{Bob}_i$'s, and Eve's systems, respectively, $P_{\text{success}}$ is the probability that a successful event occurs, $H(A|x^*,E)$ is the conditional von Neumann entropy between Alice and Eve, and $\max_{i} H(A|B_i,x^*,y_i^*)$ represents the maximal error correction cost between Alice and all Bobs. $x^*$ and $y_i^*$ correspond to Alice's and $\text{Bob}_i$'s measurement inputs for the key generation round, respectively. While calculation of the second term is relatively straightforward, calculation of the first term is challenging since we have to optimize over all Eve's eavesdropping strategies. Therefore, to calculate this term, we take two different approaches depending on the two measurement scenarios. 

For Scenario 1, we use violation of the parity-CHSH inequality~\cite{Ribeiro2018, Holz2019, Ribeiro2019}. The parity-CHSH inequality is defined as follows.
\begin{dfn}[Parity-CHSH inequality]
    Let Alice and $\text{Bob}_i$ for $i \in \{1, \ldots, N-1 \}$ be the $N$ parties of the following parity-CHSH game. Alice is asked a uniformly random binary question $x \in \{0, 1\}$ and answers a bit $a \in \{0, 1\}$. Similar to Alice, $\text{Bob}_1$ is asked a uniformly random binary question $y \in \{0, 1\}$ and answers a bit $b_1 \in \{0, 1\}$. For $i \in \{2, \ldots N-1\}$, $\text{Bob}_i$ is asked a fixed question and answers a bit $b_i \in \{0, 1\}$. Then, let $\bar{b} \equiv \oplus_{2\leq i \leq N-1} b_i$ be the parity of all the answers of $\text{Bob}_2, \ldots, \text{Bob}_{N-1}$. The parties win this game if and only if
    \begin{equation}
        a + b_1 = x (y+\bar{b}) \quad \text{mod 2}.
    \end{equation}
    The winning probability of this game $P_{\text{win}}$ with classical strategies satisfies the following inequality
    \begin{equation}
        P_{\text{win}} \leq \frac{3}{4}.
    \end{equation}
\end{dfn}
We can bound the first term in (\ref{eq:keyrate}) by using violation of the parity-CHSH inequality since the inequality is very similar to the Clauser-Horne-Shimony-Holt (CHSH) inequality~\cite{Clauser1969}. We define the following function of $P_{\text{win}}$
\begin{equation}
    f(P_{\text{win}}) \equiv h\left( \frac{1+\sqrt{(4P_\text{win}-2)^2-1}}{2} \right),
\end{equation}
where $h$ is the classical binary entropy. Then, the first term in (\ref{eq:keyrate}) is lower bounded by this function~\cite{Ribeiro2018, Grasselli2023}
\begin{equation}
    \begin{split}
        H(A|x^*,E) \geq 1- f(P_\text{win}).
    \end{split}
\end{equation}
Thus, by calculating the winning probability of the parity-CHSH game, we can get a key rate of our protocol for Scenario 1.

Next, we describe the key rate calculation for Scenario 2. We find that we cannot violate the parity-CHSH inequality with displaced single-photon detection. Therefore, we take the numerical optimization developed in Refs.~\cite{Navascues2008, Masanes2011, NietoSilleras2014, Bancal2014, Brown2024}. Let $M_{a|x}$ and $N_{b_i|y_i}$ be POVMs for Alice and Bobs and $B(H)$ be the set of bounded operators on a Hilbert space $H$. Then, we can obtain a lower bound on the conditional von Neumann entropy by solving the following optimization problem~\cite{Brown2024}
\begin{widetext}
\begin{equation}
    \begin{split}
        c_m + \inf \quad &\sum_{i=1}^{m-1} \frac{w_i}{t_i \ln 2 } \sum_{a=0}^1 \mel{\psi}{M_{a|x^*} (Z_{a,i} + Z_{a,i}^* +(1-t_i) Z_{a,i}^* Z_{a,i}) + t_i Z_{a,i} Z_{a,i}^*}{\psi}\\
        \text{subject to} \quad
        &\mel{\psi}{M_{a|x} N_{b_1|y_1} \cdots  N_{b_{N-1}|y_{N-1}}}{\psi} = P(a,b_1, \ldots, b_{N-1} | x, y_1, \ldots , y_{N-1}) \qquad \text{for all} ~ a, b_i, x, y_i\\
        &\sum_a M_{a|x} = \sum_{b_i} N_{b_i|y_i} = I \qquad \text{for all} ~ x, y_i\\
        &M_{a|x} \geq 0, \quad N_{b_i|y_i} \geq 0 \qquad \text{for all} ~ a, b_i, x, y_i\\
        & Z_{a,i}^* Z_{a,i} \leq \alpha_i^2, \quad Z_{a,i} Z_{a,i}^* \leq \alpha_i^2 \qquad \text{for all} ~ a, i\\
        &\left[M_{a|x}, N_{b_i|y_i} \right] = \left[M_{a|x}, Z_{b,i}^{(*)} \right] = \left[ N_{b_i|y_i}, Z_{a,i}^{(*)} \right] =0 \qquad \text{for all} ~ a, b_i, x, y_i, i\\
        &M_{a|x}, N_{b_i|y_i}, Z_{a,i} \in B(H) \qquad \text{for all} ~ a, b_i, x, y_i, i
    \end{split}
\label{optimization}
\end{equation}
\end{widetext}
where the infimum is taken over all quantum pure states $\ket{\psi}$ and bounded operators $M_{a|x}, N_{b_i|y_i}$, and $Z_{a,i}$, $t_1, \ldots t_m$ and $w_1, \ldots, w_m$ are the nodes and weights of an $m$-point Gauss-Radau quadrature on $[0, 1]$ with an endpoint $t_m = 1$, $c_m = \sum_{i=1}^{m-1} \frac{w_i}{t_i \ln 2}$, $\alpha_i = \frac{3}{2} \max \{ \frac{1}{t_i}, \frac{1}{1-t_i} \}$, and $P(a, b_1, \ldots, b_{N-1} | x, y_1, \ldots, y_{N-1})$ is the observed probability distribution in the Bell test rounds. Since we do not have any efficient methods to solve the above optimization problem, we relax this problem into a semidefinite program (SDP) by using Navascu{\'{e}}s-Pironio-Ac{\'{i}}n (NPA) hierarchy~\cite{Navascues2008} (see Appendix\ref{appendixd}). We perform the NPA hierarchy using NCPOL2SDPA~\cite{Wittek2015} and solve SDPs using MOSEK~\cite{Mosek2024}.

Calculation of the success probability $P_\text{success}$ and the classical entropy $\max_i H(A|B_i, x^*, y_i^*)$ is straightforward when we obtain explicit form of joint quantum states. We show how to calculate the joint quantum states in Appendix~\ref{appendixb}.

\begin{figure}[tbp]
 \centering
 \includegraphics[keepaspectratio, scale=0.45]{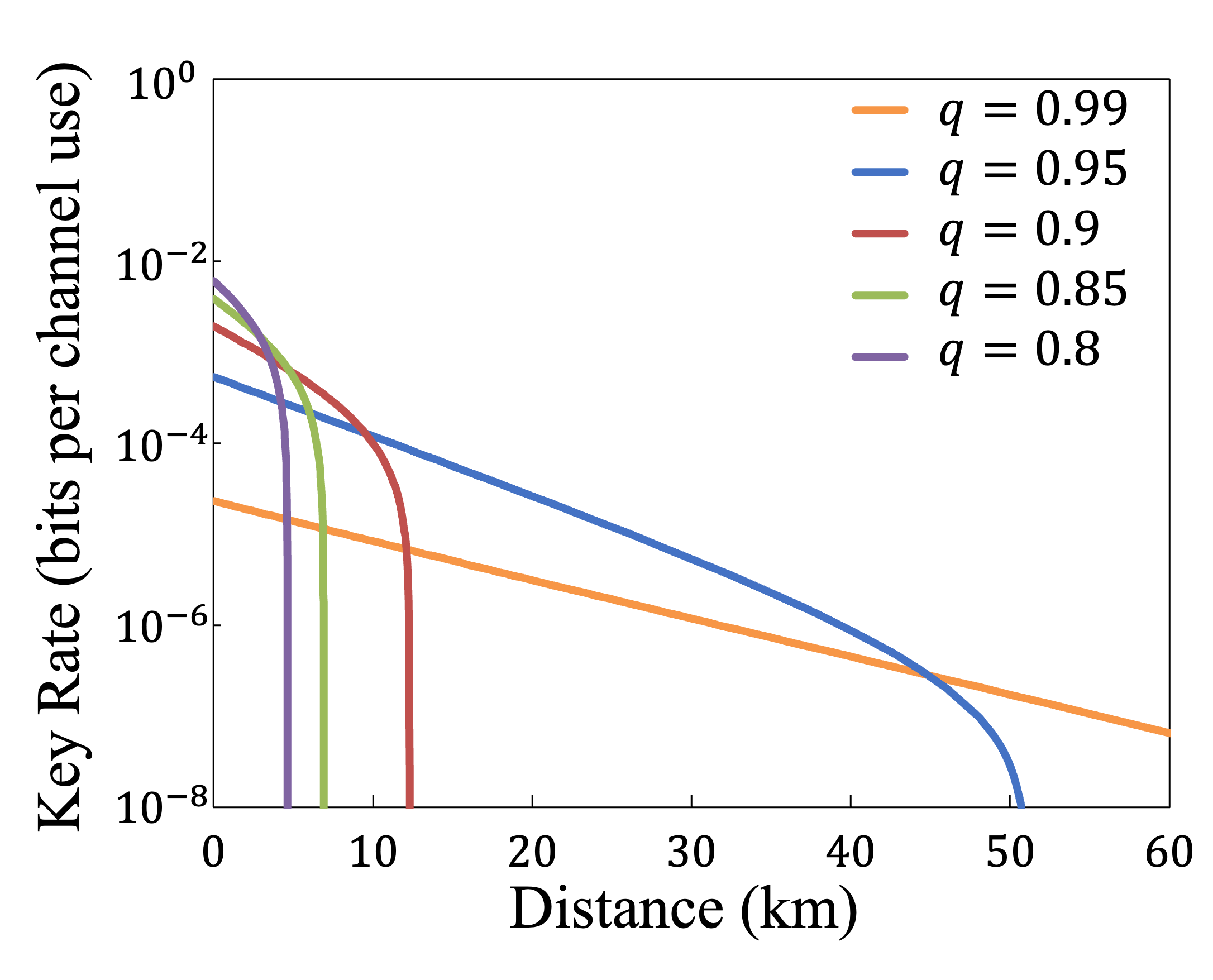}
 \caption{Key rate versus distance $L$ for different $q$ for Scenario 1, where $N=4$. We take $\eta_e = 0.97$ and $p_{DC} = 10^{-6}$.}
 \label{fig:ParityQ4}
\end{figure}

\begin{figure*}[htbp]
 \centering
 \includegraphics[keepaspectratio, scale=0.55]{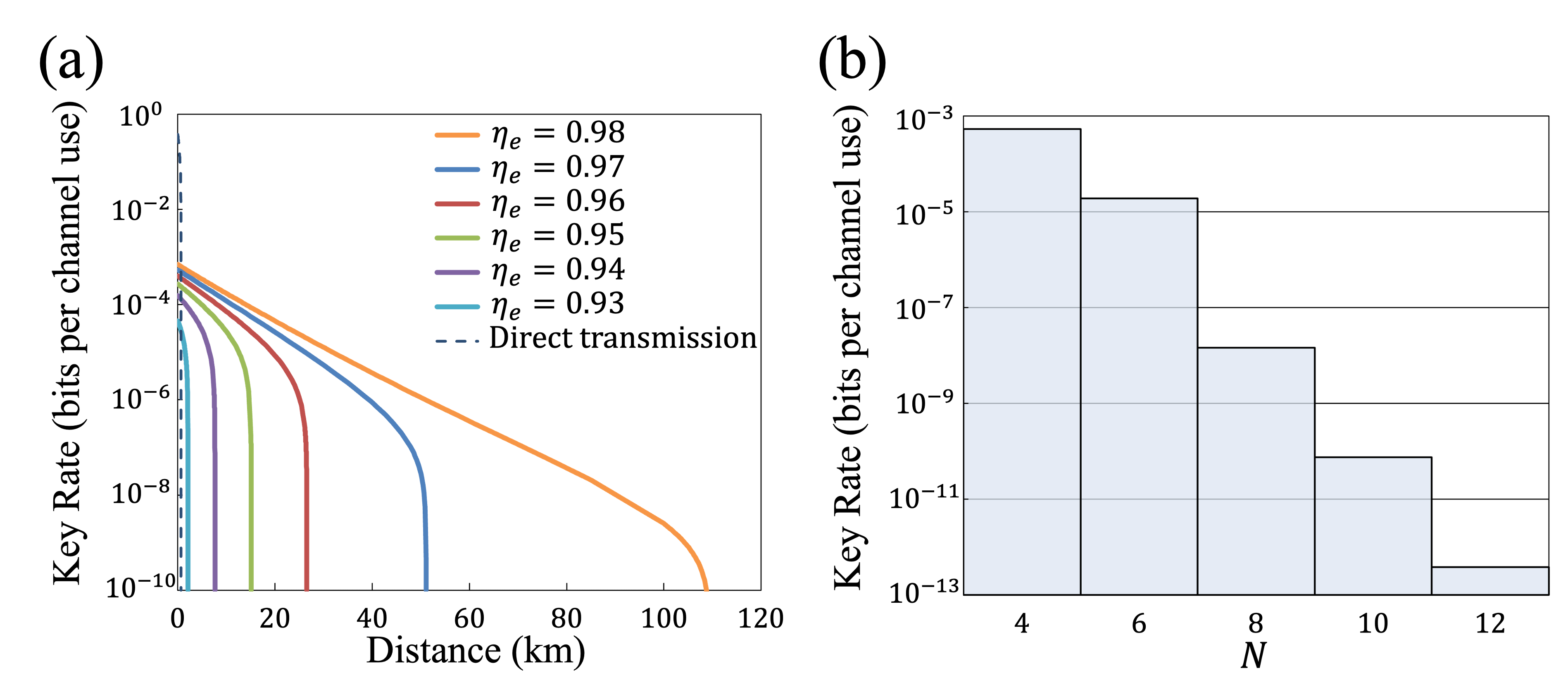}
 \caption{(a) Key rate versus distance $L$ for different detection efficiencies $\eta_e$ for Scenario 1, where $N = 4$. Black dashed line shows key rates of the direct transmission protocols with a detection efficiency of 97 \% and dark count probability of $10^{-6}$. (b) Key rate versus the number of parties $N$ where $L=0$ and $\eta_e = 0.97$. For both plots, we take $q = 0.95$ and $p_{DC} = 10^{-6}$.}
 \label{fig:ParityDifferentEtae}
\end{figure*}

\section{RESULTS AND DISCUSSION}\label{section:results}
Here, we show the results of our key rate calculation. We consider channel loss of 0.2 dB/km corresponding to a transmittance of $\eta = 10^{-0.02L}$, where $L$ is the distance in kilometers between the central station and each party's site. For simplicity, we assume that dark count probabilities of the detectors at the central station and each party's site are the same; $p_{DC}^e = p_{DC}$.

\textit{Key rates for Scenario 1}. In Fig.~\ref{fig:ParityQ4}, we show key rates of our protocol for four parties with different values of $q$ in (\ref{initial}). Here, we fix $\eta_e = 0.97$ and $p_{DC} = 10^{-6}$. We find a relationship between $q$ and key rates. When $q$ is small, the key rate at short distance is large, but the maximum distance for secure key generation is small. On the other hand, when $q$ is large, the key rate is small at short distance but we can distribute a key over long distance. This trade-off can be interpreted as follows. When $q$ is small, the probability that each party sends a single-photon to the central station is large, and thus the success probability $P_{\text{success}}$ is high. However, large probability of sending single-photons means that there are undesired multiphoton events on the distributed GHZ states. For instance, when three of four legitimate parties send a single-photon and one single-photon is lost during channel transmission, this event is counted as the successful event. Then, a distributed state over the legitimate parties is not the desired GHZ state. Therefore, fidelity of a distributed state decreases due to such multiphoton events. This makes a key rate large at short distance but small at long distance. On the other hand, if $q$ is large, the success probability will decrease since the probability each party sends a single-photon is small. However, in this situation, the effect of the undesired multiphoton events is relatively small. Thus, we can observe the trade-off between $q$ and key rate. From now on, we calculate key rates for $q = 0.95$ since this value balances high key rates and long communication distances.

In Fig.~\ref{fig:ParityDifferentEtae} (a), we plot the key rates of our protocol for $N=4$ with different values of detection efficiency $\eta_e$. We fix the input state coefficient in (\ref{initial}) to $q = 0.95$ and $p_{DC} = 10^{-6}$. We also plot the key rates of a direct transmission protocol where a GHZ state is locally generated and distributed over legitimate parties, see Appendix~\ref{appendixc}. For the direct transmission protocol, we consider a detection efficiency of 97 \% and dark count probability of $10^{-6}$. While the direct transmission protocol distributes a secret key over few hundreds meters, our protocol can distribute a secret key over tens of kilometers. We find that the detection efficiency $\eta_e$ has a significant effect on key rates and we require higher detection efficiency when the number of parties increases. Also, in Fig.~\ref{fig:ParityDifferentEtae} (b), we plot key rates against the number of parties $N$ where $L =0$ and $\eta_e = 0.97$. It can be observed that the key rates exponentially decrease with increasing $N$. This decrease occurs since all copies of the four-party-interferometer must have the successful detection pattern simultaneously for $N \geq 6$.

\begin{figure}[tbp]
 \centering
 \includegraphics[keepaspectratio, scale=0.45]{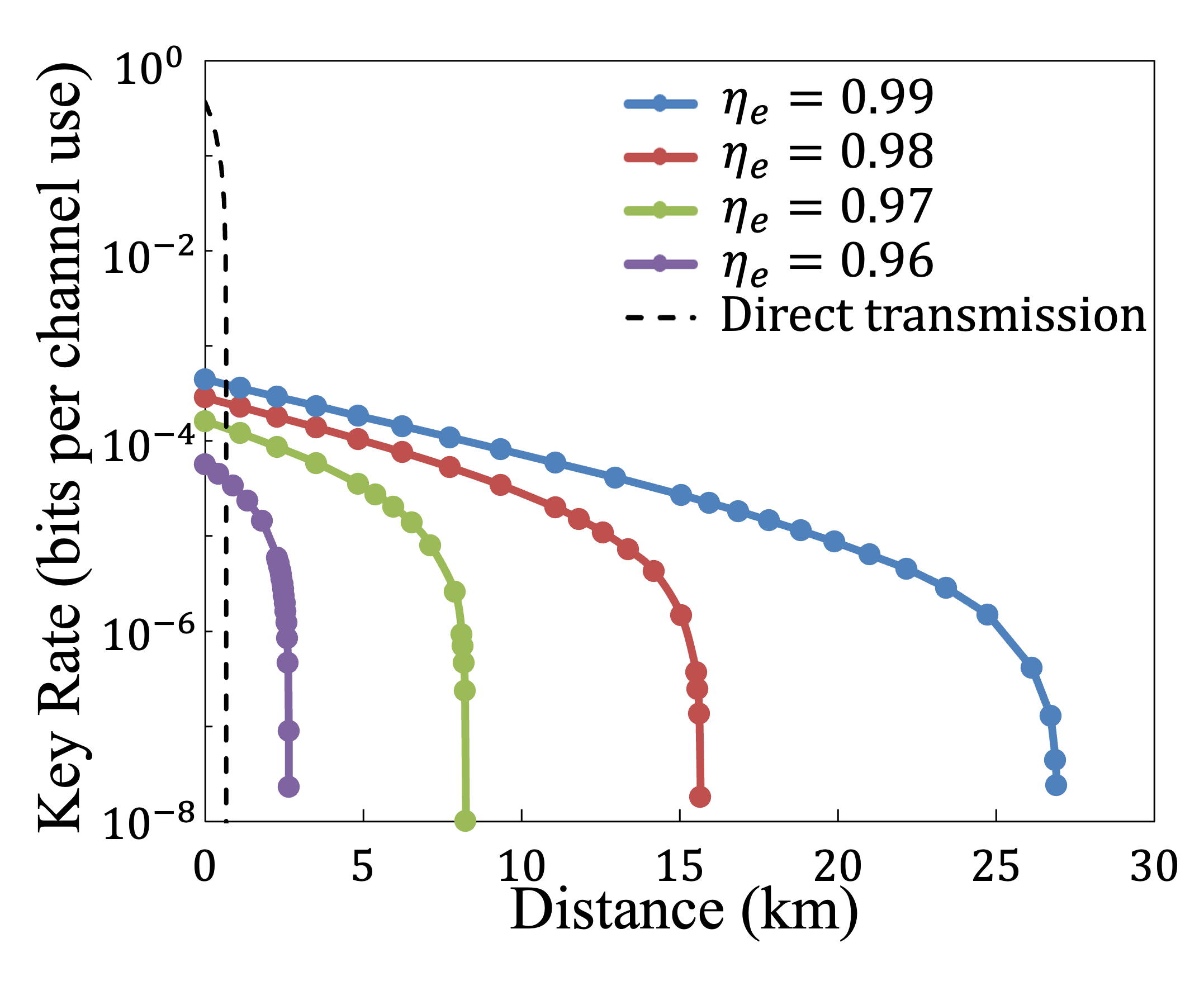}
 \caption{Key rate versus distance $L$ for different detection efficiencies $\eta_e$ for Scenario 2, where $N=4$. We take $q = 0.95$ and $p_{DC} = 10^{-6}$. Black solid line is key rate of the direct transmission protocol with a detection efficiency of 97 \% and dark count probability of $10^{-6}$.}
 \label{fig:QuasiEtae4}
\end{figure}

\textit{Key rates for Scenario 2}. In Fig.~\ref{fig:QuasiEtae4}, we plot key rates of our protocol with different values of detection efficiency $\eta_e$ where the number of parties $N=4$ and the number of the nodes in the Gauss-Radau quadrature $m=4$. We take $q = 0.95$ and $p_{DC} = 10^{-6}$ for the key rates. We also show a key rate of the direct transmission protocol with a detection efficiency of 97 \% and dark count probability of $10^{-6}$. We find that our protocol can overcome the direct transmission protocol even with the feasible measurement. We also find that minimum detection efficiency to obtain a secret key is 96\% and this value is higher than that of Scenario 1.

\begin{figure}[htbp]
 \centering
 \includegraphics[keepaspectratio, scale=0.45]{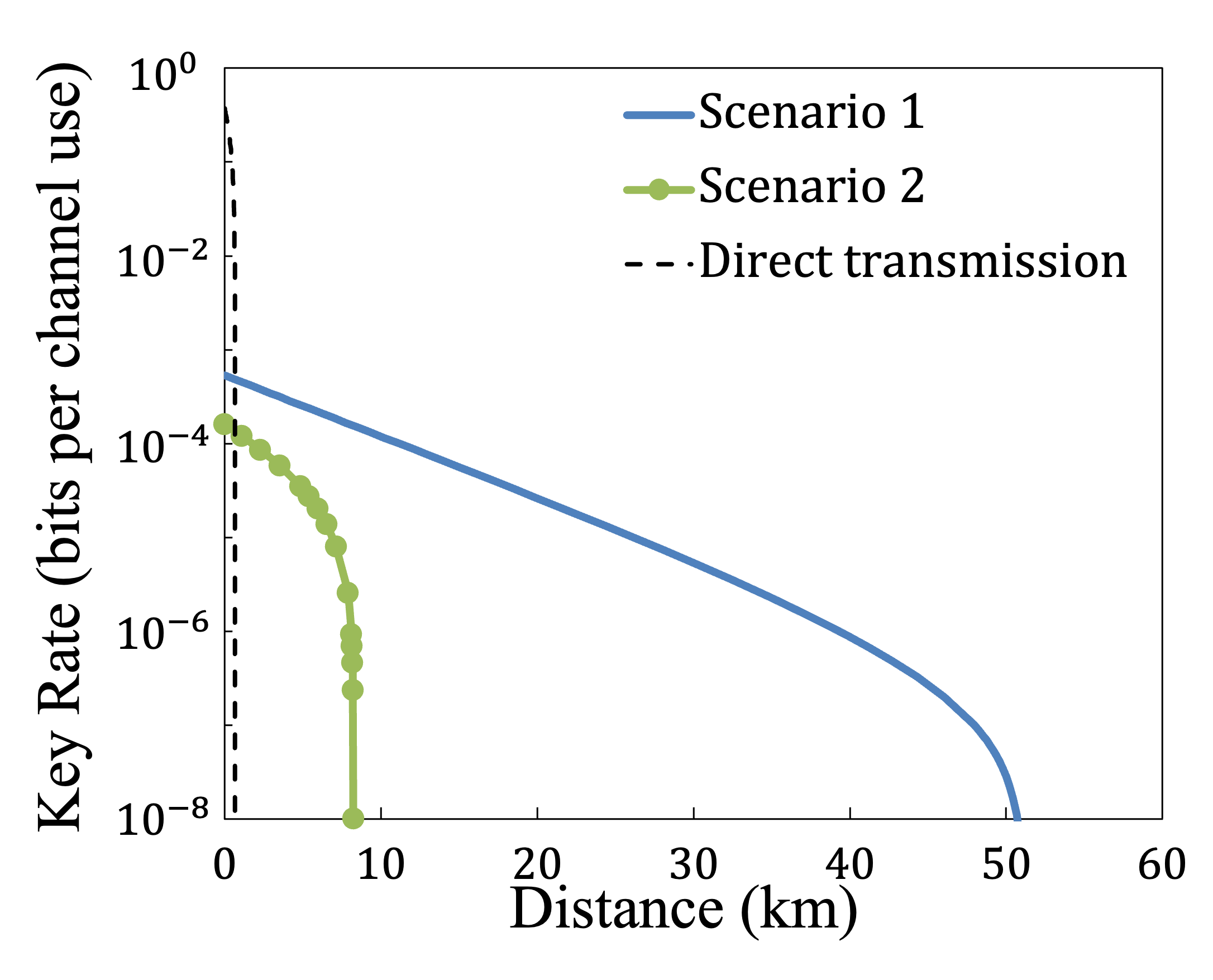}
 \caption{Key rate versus distance $L$ where $N=4$. Blue solid line is key rates of Scenario 1. Green circles are key rates of Scenario 2. We take $q = 0.95$, $\eta_e = 0.97$ and $p_{DC} = 10^{-6}$. Black dashed line is a key rate of the direct transmission protocol with a detection efficiency of 97 \% and dark count probability of $10^{-6}$.}
 \label{fig:Comparison}
\end{figure}

We compare key rates for Scenario 1 and Scenario 2 where $N=4$ in Fig.~\ref{fig:Comparison}. The blue solid line shows key rates of Scenario 1. The green circle shows a key rate of Scenario 2. For these plots, we take $q = 0.95$, $\eta_e = 0.97$ and $p_{DC} = 10^{-6}$. We also plot the key rate of the direct transmission protocol with a detection efficiency of 97 \% and dark count probability of $10^{-6}$. We find that the maximal communication distance of Scenario 1 is more than five times longer than that of Scenario 2. This difference comes from the fact that we cannot perform arbitrary Pauli measurements with displacement and photon detection.

Finally, we discuss requirement of detection efficiency $\eta_e$ of our protocol. For $N=4$, we find that minimum detection efficiency to obtain a secret key of our protocol is 93\% for Scenario 1 and 96\% for Scenario 2. These values can be achieved with state-of-the-art detectors~\cite{Reddy2020, Chang2021, Liu2022}. However, it is still a challenge to fulfill this requirement for a whole system. There are several theoretical works to relax the required detection efficiency of DI-QKD protocols~\cite{Ma2012, Tan2020, Ho2020, Xu2022}. Therefore, it is an important future work that we apply these techniques to our protocol and show the experimental feasibility of our protocol with current technology.



\section{CONCLUSION}\label{section:conclusion}
In this paper, we propose a long-distance DI-CKA protocol which is based on the efficient distribution of GHZ states. We consider two different measurement scenarios, that is, arbitrary measurement scenario and displacement operation and photon detection scenario. For the former scenario, we calculate key rates of our protocol using violation of the parity-CHSH inequality. For the latter scenario, we calculate key rates by using the numerical optimization approach. We compare our key rates with a key rate of the direct transmission protocol where a GHZ state is locally generated and distributed over legitimate parties. Then, we find that our protocol overcomes the direct transmission protocol in terms of the distance for both of the measurement scenarios. We analyze the required detection efficiency to obtain a secret key. We also analyze the effect of the parameter $q$ of the single-photon entanglement in (\ref{initial}) on the key rate. Although we consider asymptotic key rates in this paper, understanding the effects of finite statistics is important for experimental realization. Analyzing the consequences of finite statistics is beyond the scope of this paper, and we leave such an analysis for future research.

\begin{acknowledgments}
This work was supported by JST SPRING, Grant No. JPMJSP2123, JST Moonshot R\&D, Grant No. JPMJMS226C and Grant No. JPMJMS2061, JST CRONOS, Grant No. JPMJCS24N6, and JST ASPIRE, Grant No. JPMJAP2427. We also acknowledge support from the Danish National Research Foundation, Center for Macroscopic Quantum States (bigQ, DNRF142), the  European Union’s Horizon Europe research and innovation programme under the project ``Quantum Security Networks Partnership'' (QSNP, grant agreement no. 101114043), and from Innovation Fund Denmark (CyberQ, grant agreement no. 3200-00035B).

\end{acknowledgments}

\appendix
\section{INTERFEROMETER AT THE CENTRAL STATION}\label{appendixA}
Here, we describe details of the interferometer at the central station in our protocol. The interferometer is composed of 50:50 beamsplitters and single-photon detectors. Operation of the interferometer on creation operators corresponding each modes can be expressed as
\begin{equation}
\begin{split}
    U = u^{\otimes 2},
\end{split}
\end{equation}
where 
\begin{equation}
    u = \frac{1}{\sqrt{2}} \left( 
    \begin{array}{cc}
        1 & 1 \\
        -1 & 1
    \end{array}
    \right).
\end{equation}
Let $\boldsymbol{a} = \{a^\dagger_1, a^\dagger_2, a^\dagger_3, a^\dagger_4\}^T$ be creation operators of each mode before the interferometer and $\boldsymbol{b} = \{b^\dagger_1, b^\dagger_2, b^\dagger_3, b^\dagger_4\}^T$ be creation operators of each mode after the interferometer. Then, we can express $\boldsymbol{b}$ by using $\boldsymbol{a}$ and $U$;
\begin{equation}\label{interferometertransformation}
\begin{split}
    \boldsymbol{b}
    &= U \boldsymbol{a} \\
    &= \frac{1}{2} \left( \begin{array}{c}
         a^\dagger_1 + a^\dagger_2 + a^\dagger_3 + a^\dagger_4  \\
         a^\dagger_1 - a^\dagger_2 + a^\dagger_3 - a^\dagger_4  \\ 
         a^\dagger_1 + a^\dagger_2 - a^\dagger_3 - a^\dagger_4  \\
         a^\dagger_1 - a^\dagger_2 - a^\dagger_3 + a^\dagger_4  \\
    \end{array}
    \right).
\end{split}
\end{equation}
Here, when detectors $D_1$ and $D_2$ detect a single-photon, possible inputs for the interferometer are
\begin{equation}
\begin{split}
    b^\dagger_1 b^\dagger_2 &= \frac{1}{4} \left( a^\dagger_1 + a^\dagger_2 + a^\dagger_3 + a^\dagger_4 \right) \left( a^\dagger_1 - a^\dagger_2 + a^\dagger_3 - a^\dagger_4 \right)\\
    &= \frac{1}{4} ( (a^\dagger_1)^2 - (a^\dagger_2)^2 + (a^\dagger_3)^2 - (a^\dagger_4)^2\\
    &\quad +2 a^\dagger_1 a^\dagger_3 - 2 a^\dagger_2 a^\dagger_4 ).
\end{split}
\end{equation}
Since each party does not send more than one photon, clicks of detectors $D_1$ and $D_2$ mean that the input of the interferometer was the GHZ state in the following form
\begin{equation}
\begin{split}
    b^\dagger_1 b^\dagger_2 &= \frac{1}{2}  \left(a^\dagger_1 a^\dagger_3 - a^\dagger_2 a^\dagger_4 \right) \\
    &= \frac{1}{2} \left( \ket{1010} - \ket{0101} \right).
\end{split}
\end{equation}
In our protocol, since all the legitimate parties send the single-photon entanglement defined in (\ref{initial}), a quantum state distributed over the parties is equivalent to the above GHZ state up to a global phase. The number of click patterns where two of four detectors click is 6 and distributed GHZ states which correspond to each click pattern are shown in Tab.~\ref{Tab:DistributedGHZstate}. All the GHZ states contain only two photons and this means that we can distribute those states with a rate which scales $\mathcal{O}\left( \eta^2 \right)$.

\begin{figure}[tbp]
\includegraphics[keepaspectratio, scale=0.35]{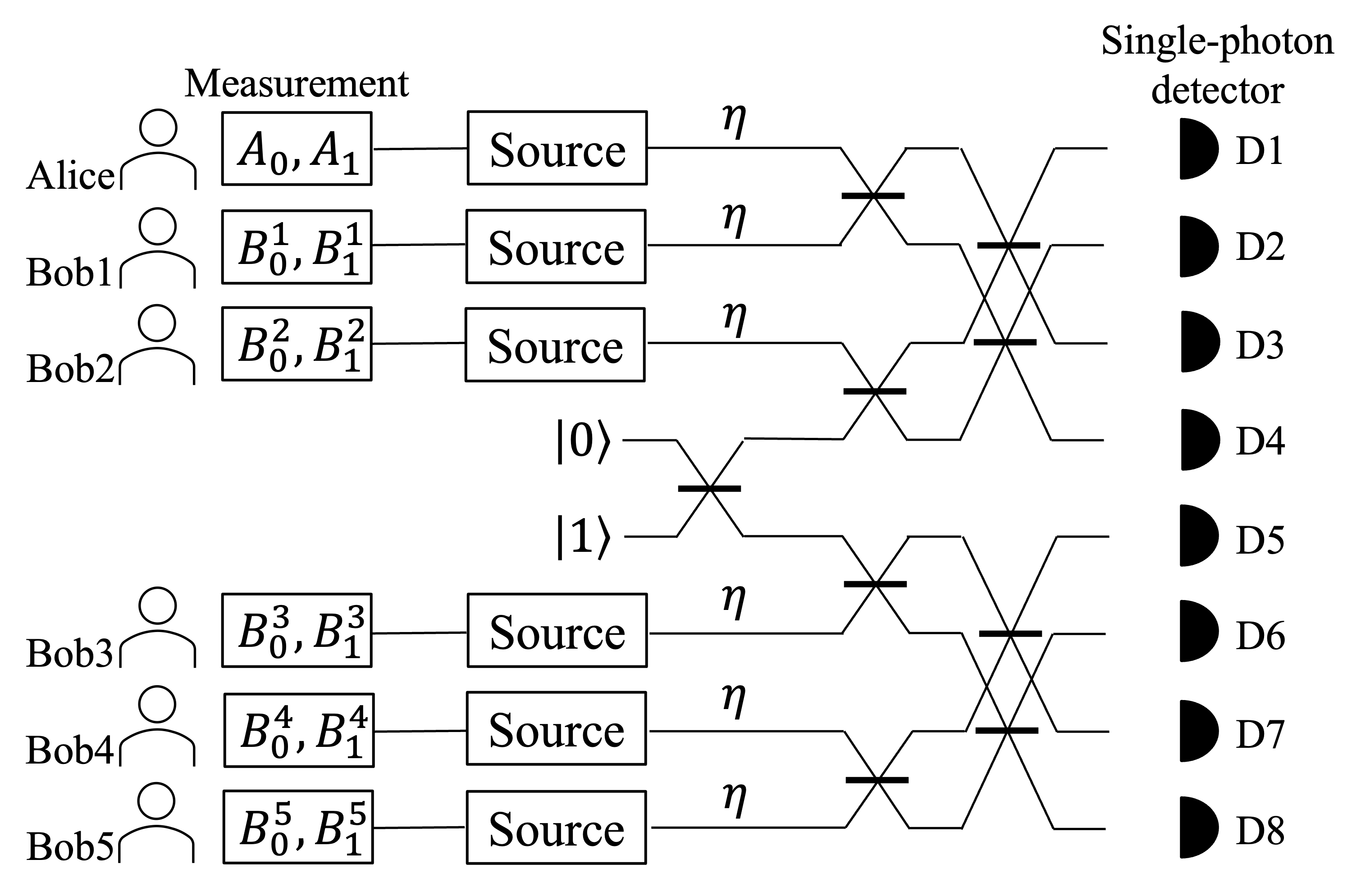}
\caption{\label{fig:DICKASetup6}
Schematic of our DI-CKA protocol for $N=6$. We prepare two copies of the interferometer for four modes and input single-photon entanglement into auxiliary modes between the interferometers. The legitimate parties distil a secret key from events where the single-photon detectors $D_1$, $D_2$, $D_5$ and $D_6$ detect a single-photon. Rest of the protocol is the same as that for four parties.}
\end{figure}

Next, let us explain an extension of our protocol to more than four parties. We show a schematic of our protocol for six parties in Fig.~\ref{fig:DICKASetup6}. Suppose we have two separate GHZ states for four modes $(\ket{1010}-\ket{0101})/\sqrt{2}$. Performing Bell state measurement between the two GHZ states, we can get the GHZ state for six parties: $(\ket{101010} - \ket{010101})/2\sqrt{2}$. Therefore, we can extend our protocol to a six-part protocol. Specifically, we prepare two copies of the interferometer and input a single-photon entangled state into two auxiliary modes as shown in Fig.~\ref{fig:DICKASetup6}. Here, the events where the detectors $D_1$, $D_2$, $D_5$ and $D_6$ detect a single-photon are considered as the successful events. Extension to more than six parties is straightforward \cite{Shimizu2025}. When the number of legitimate parties is $N$, we prepare $N/2-1$ copies of the interferometer and input single-photon entanglement into auxiliary modes between each interferometers.

\section{NPA HIERARCHY}\label{appendixd}
We describe the NPA hierarchy that we use for key rate calculation for Scenario 2. Full details can be found in the original paper \cite{Navascues2008}. 

By solving the optimization problem in (\ref{optimization}), we get a lower bound on the conditional entropy in (\ref{eq:keyrate}). However, we do not have efficient methods to solve this problem. Thus, we use the NPA hierarchy to relax this problem into SDP which can be efficiently solved.

Let $\ket{\psi}$ be a quantum state on a Hilbert space $H$. We define a set of Alice's measurement operators $\mathcal{A} = \{ M_{a|x} \}$ on $H$. We analogously define a set of $\text{Bob}_i$'s measurement operators $\mathcal{B}_i$ ($i \in \{ 1, \ldots, N-1 \}$) and a set of Eve's measurment operators $\mathcal{Z}$. Then, let $\mathcal{E}$ denote a set of these operators plus the identity $I$, that is, $\mathcal{E} = I \cup \mathcal{A} \cup \mathcal{B}_1 \cup \cdots \cup \mathcal{B}_{N-1} \cup \mathcal{Z}$. Suppose we minimize a linear combination of expectation values $\ev{O_i^\dagger O_j} = \text{Tr} \left[ \ketbra{\psi}{\psi} O_i^\dagger O_j \right]$ where $O_i ~ (i \in \{1, \ldots, n \})$ is a linear combination of products of operators in $\mathcal{E}$. The objective function in (\ref{optimization}) is expressed in this form. We can relax this problem into SDP in the following way. In the $k$-th level NPA hierarchy, we construct a $k$-th moment matrix $\Gamma^{(k)}$ (also referred as Gram matrix) whose entry is $\Gamma_{ij}^{(k)} = \ev{(O_i^k)^\dagger O_j^k}$ where $O_i^k$ denotes a product of at most $k$ operators in $\mathcal{E}$. It is known that matrices constructed in this way are positive semidefinite. Also, the objective function is expressed as a linear combination of entries of the moment matrix. Therefore, the original optimization problem now can be interpreted as SDP. Note that $k$-th moment matrix does not necessarily correspond to feasible points of an original optimization problem, that is, the range of optimization in the NPA hierarchy is wider than that of the original optimization. However, in the limit of $k \to \infty$, it is proved that a feasible region of the original optimization is equivalent to that of SDP constructed from the NPA hierarchy.

\section{CALCULATION OF DISTRIBUTED QUANTUM STATE}\label{appendixb}
We calculate distributed quantum states over legitimate parties considering channel losses, detector efficiencies and dark counts of detectors at the central station and each party's site. We can model the effect of the detection efficiency as an action of a beamsplitter with vacuum in one of the input modes. Here, we discuss $N=4$, and extension to more than four parties is straightforward.  For simplicity, we assume that $\eta_d = 1$. Each party prepares a single-photon entanglement $\sqrt{q} \ket{10}_{XX'} + \sqrt{1-q} \ket{01}_{XX'}$, where mode $X'$ is transmitted to the central station and mode $X$ is detected by the party. The quantum state after applying the beamsplitter operations corresponding to the channel loss and the detection efficiency to this single-photon entanglement is as follows
\begin{equation}
    \begin{split}
        \ket{S}_{XEF}\ket{1}_{X'} + \ket{V}_{XEF} \ket{0}_{X'},
    \end{split}
\end{equation}
where
\begin{equation}
\begin{split}
    \ket{S} &= \sqrt{1-q} \sqrt{\eta} \ket{000}_{XEF},\\
    \ket{V} &= \sqrt{q} \left( \sqrt{\eta_e} \ket{100}_{XEF} + \sqrt{1-\eta_e} \ket{001}_{XEF} \right)\\
    &\quad+ \sqrt{1-q} \sqrt{1-\eta} \ket{010}_{XEF}.
\end{split}
\end{equation}
Here, $E$ denotes an environmental system of channel transmission and $F$ denotes an environmental system of each party's detector.

Now, we focus on the detection part of the interferometer. When dark count probability of a single-photon detector is $p_{DC}$, we can express its measurement as follows
\begin{equation}
    \{ (1-p_{DC}) \ketbra{0}, \ketbra{1} + p_{DC} \ketbra{0}  \}.
\end{equation}
Here, we assume that the single-photon detectors at the central station are photon-number resolving and we discard events where more than 1 photon arrive at each detector. Then, the click pattern where only $D_1$ and $D_2$ detect a single-photon corresponds to
\begin{equation}
    \begin{split}
        &\left( \ketbra{1} + p_{DC} \ketbra{0} \right)_{D_1} \left( \ketbra{1} + p_{DC} \ketbra{0} \right)_{D_2}\\
        &\quad \otimes \left( (1-p_{DC}) \ketbra{0} \right)_{D_3}\left( (1-p_{DC}) \ketbra{0} \right)_{D_4}\\
        = &\left(1-p_{DC}\right)^2 \ketbra{1}\ketbra{1}\ketbra{0}\ketbra{0}\\
        + &\left(1-p_{DC}\right)^2p_{DC} \ketbra{1}\ketbra{0}\ketbra{0}\ketbra{0}\\
        + &\left(1-p_{DC}\right)^2 p_{DC} \ketbra{0}\ketbra{1}\ketbra{0}\ketbra{0}\\
        + & \left( 1-p_{DC} \right)^2 p_{DC}^2 \ketbra{0}\ketbra{0}\ketbra{0}\ketbra{0}
    \end{split}
\end{equation}
We find that there are four situations corresponding to this click pattern: 
\begin{enumerate}
    \item no dark counts,
    \item $D_1$ detects a single-photon and a dark count occurs at $D_2$,
    \item $D_2$ detects a single-photon and a dark count occurs at $D_1$,
    \item dark counts occur at both $D_1$ and $D_2$.
\end{enumerate}
From the transformation matrix of the interferometer in (\ref{interferometertransformation}), a heralded quantum state corresponding to the first situation is
\begin{equation}
    \ket{\phi^1} = \frac{1}{2}(\ket{S}\ket{V}\ket{S}\ket{V} -\ket{V}\ket{S}\ket{V}\ket{S}).
\end{equation}
Analogously, heralded quantum states corresponding to the second, third and fourth situation can be written as follows
\begin{equation}
    \begin{split}
        \ket{\phi^2} &= \frac{1}{2} (\ket{S}\ket{V}\ket{V}\ket{V} + \ket{V}\ket{S}\ket{V}\ket{V}\\ &\quad + \ket{V}\ket{V}\ket{S}\ket{V} + \ket{V}\ket{V}\ket{V}\ket{S}),\\
    \end{split}
\end{equation}
\begin{equation}
    \begin{split}
        \ket{\phi^3} &= \frac{1}{2} (\ket{S}\ket{V}\ket{V}\ket{V} - \ket{V}\ket{S}\ket{V}\ket{V}\\ &\quad  + \ket{V}\ket{V}\ket{S}\ket{V} - \ket{V}\ket{V}\ket{V}\ket{S}),\\      
    \end{split}
\end{equation}
\begin{equation}
    \ket{\phi^4} = \ket{V}\ket{V}\ket{V}\ket{V}.  
\end{equation}
Thus, a quantum state distributed over legitimate four parties is
\begin{equation}
    \rho_X = \text{Tr}_{EF} \left[ \sigma_{XEF} \right],
\end{equation}
where
\begin{equation}
    \begin{split}
        \sigma_{XEF} &= \frac{1}{P} (P_1 \ketbra{\phi^1}{\phi^1} + P_2 \ketbra{\phi^2}{\phi^2}\\
        &\quad+ P_3\ketbra{\phi^3}{\phi^3} + P_4 \ketbra{\phi^4}{\phi^4}),\\
        P_1 &= (1-p_{DC})^2, \\
        P_2 &= (1-p_{DC})^2 p_{DC}, \\
        P_3 &= (1-p_{DC})^2 p_{DC}, \\
        P_4 &= (1-p_{DC})^2 p_{DC}^2, \\
        P &= \text{Tr} [ P_1 \ketbra{\phi^1}{\phi^1} + P_2 \ketbra{\phi^2}{\phi^2} \\
        &\quad +P_3 \ketbra{\phi^3}{\phi^3} + P_4 \ketbra{\phi^4}{\phi^4} ].\\ 
    \end{split}
\end{equation}
Then, the success probability $P_\text{success}$ is given as
\begin{equation}
    P_\text{success} = P.
\end{equation}

For $N=6$, we can calculate a four-mode heralded quantum state conditioned on the clicks of the single-photon detectors $D_1$ and $D_2$ in the same way as $N = 4$. Also, we can calculate a four-mode heralded quantum state where the detectors $D_5$ and $D_6$ detect a single-photon. Then, we get a joint quantum state distributed over the legitimate parties by projecting two auxiliary modes onto the Bell state and tracing out environmental modes. The success probability for $N = 6$ is expressed as a product of the probability that the detectors $D_1$ and $D_2$ detect a single-photon, the probability that the detectors $D_5$ and $D_6$ detect a single-photon and the probability that the projection onto the Bell state succeeds.

\section{KEY RATE OF DIRECT TRANSMISSION PROTOCOL}\label{appendixc}
In this section, we describe the direct transmission protocol and its key rate calculation~\cite{Ribeiro2019}. In the direct transmission protocol, a $N$-partite GHZ state, $(\ket{00\cdots 0} + \ket{11 \cdots 1})/\sqrt{2}$, is generated locally and distributed over $N$ legitimate parties. Each party performs measurements on the GHZ state and obtains a secret key the same as in our protocol. Here, we model each party's measurement in the following way. Let $\Pi (\theta) = \cos{\theta} \sigma_Z + \sin{\theta} \sigma_X$ and $M^d(\theta) = M_0^d (\theta) - M_1^d (\theta)$, where
\begin{equation}
    \begin{split}
        M_0^d (\theta) &= \{1-(1-p_{DC})(1-\eta) \} \frac{I+\Pi(\theta)}{2}\\
        &\quad+ p_{DC} \frac{I-\Pi(\theta)}{2},\\
        M_1^d (\theta) &= I-M_0^d,
    \end{split}
\end{equation}
$\eta$ is detection efficiency and $p_{DC}$ is dark count probability. Then, for the key generation rounds, every party performs $M^d(0)$. For the Bell test rounds, Alice performs $A_0 = M^d(0)$ or $A_1 = M^d (\pi/2)$, $\text{Bob}_1$ performs $B_0^1 = M^d(\pi/4)$ or $B_1^1 = M^d(-\pi/4)$, and the other Bobs perform $B_0^i = B_1^i = M^d(\pi/2)$ for $i \in \{2, \ldots , N-1 \}$. A key rate of the direct transmission protocol is calculated by violation of the parity-CHSH inequality same as our protocol for the arbitrary measurement scenario.

\bibliography{DIQKD}

\end{document}